\documentclass{PoS}

\usepackage{amsmath}
\usepackage{graphicx,comment}

\def\Re {\mathop{\hbox{Re}}}

\def\Tr {\mathop{\hbox{Tr}}}

\newcommand{\beq}{\begin{equation}}
\newcommand{\eeq}{\end{equation}}

\title{Winding number expansion for the canonical approach to finite density simulations}

\ShortTitle{Winding number expansion in canonical approach to finite
density }
\author{\speaker{Xiangfei Meng}\\
       Department of Physics and Astronomy, University of
Kentucky, Lexington KY 40506, USA\\
Department of Physics, Nankai University, Tianjin 300071, China\\
       E-mail: \email{mengxf@mail.nankai.edu.cn}}
\author{Anyi Li\\
       Department of Physics and Astronomy, University of
Kentucky, Lexington KY 40506, USA\\
       E-mail: \email{anyili@pa.uky.edu}
       }

\author{Andrei Alexandru\\
       Physics Department, The George Washington University, Washington, DC 20052, USA\\
       E-mail: \email{aalexan@gwu.edu}
       }
\author{Keh-Fei Liu\\
       Department of Physics and Astronomy, University of
Kentucky, Lexington KY 40506, USA \\
E-mail: \email{liu@pa.uky.edu}
       }

\abstract{ The canonical partition function approach was designed to
avoid the overlap problem that affects the lattice simulations of
nuclear matter at high density. The method employs the projections
of the quark determinant on a fix quark number sector. When the
quark number is large, the evaluation of the projected determinant
becomes numerically unstable. In this paper a different evaluation
method based on expanding the determinant in terms of loops winding
around the lattice is studied. We show that this method is stable
and significantly faster than our original algorithm. This greatly
expands the range of quark numbers that we can simulate
effectively.}

\FullConference{The XXVI International Symposium on Lattice Field Theory \\
July 14 - 19, 2008\\
Williamsburg, Virginia, USA}

\begin{document}

\section{Introduction}
To simulate QCD at finite density, the usual approach based on the
grand canonical partition function is to split the fermionic
determinant into two parts: a real and positive part used to
generate an ensemble of configurations, and a complex phase which
is folded into the observables. This approach has two major
drawbacks: the sign problem and the overlap problem. In order to
address the overlap problem, a method which employs the canonical
partition function has been proposed~\cite{kfl05}.

To generate ensembles using this method, the projected
determinant for a fixed net quark number (number of quarks
minus the number of anti-quarks) needs to be calculated using the
Fourier transform of the fermionic determinant~\cite{afhl05, la07}.
To evaluate it, a discrete version of the Fourier transform was used~\cite{afhl05}
which was shown to be accurate~\cite{la07} for quark numbers as large at 15.
However, when simulating larger lattices, to achieve the same densities we need
to use a proportionally larger number of quarks. As it turns out, the
discrete Fourier transform becomes unstable when we use quark numbers larger than
20. To overcome this problem, we have developed a winding number expansion method (WNEM),
a method based on the expansion in terms of quark loops. Using this approach,
we can evaluate reliably the projected determinant for large quark
numbers.

\section{Canonical approach}

The canonical partition function for lattice QCD is related to the
grand-canonical partition function via the fugacity expansion
\begin{equation}
Z(V,T,\mu)=\sum_{k}Z_{c}(V,T,k)e^{\mu k/T},
\end{equation}
where $k$ is the net number of quarks and $Z_{c}$ is the canonical
partition function. Thus, the canonical
partition function can be written as a Fourier transform of the
grand-canonical partition function
\begin{equation}
Z_{c}(V,T,k)=\int_{0}^{2\pi}\frac{d\phi}{2\pi}e^{-ik\phi}Z(V,T,\mu)|_{\mu=i\phi T}.
\end{equation}
In this paper, we will consider only the case of two degenerate flavors. If we denote
$M$ as the quark matrix for one quark flavor, we have
\begin{eqnarray}
\label{mainidea}
Z_{c}(V,T,k) &=&\int \mathcal{D}U e^{-S_{G}[U]}\det{_{k}}M^{2}[U] \\
{\nonumber}
&=&\int \mathcal{D}U
e^{-S_{G}[U]}\det M^{2}[U]\frac{|\Re\det{_{k}}M^{2}[U]|}{\det
M^{2}[U]}\frac{\det{_{k}}M^{2}[U]}{|\Re\det{_{k}}M^{2}[U]|},
\end{eqnarray}
where
\begin{equation}
\det{_{k}} M^{2}[U]=\int_{0}^{2\pi}\frac{d\phi}{2\pi}e^{-ik\phi} \det M^{2}[U,\phi]
\label{discreteFT}
\end{equation}
is the projected determinant with a fixed net quark number $k$.

\section{Discrete Fourier transform instability}

The simplest way to compute the Fourier transform, Eq.~(\ref{discreteFT}),
is to replace the continuous transform with its discrete version,
\begin{equation}
\widetilde{\det{_{k}}}M^{2}[U]=\frac{1}{N}\sum_{j=0}^{N-1}e^{-ik\phi_{j}}
\det M^{2}[U,\phi_{j}]{\hspace{1cm}}\phi_{j}=\frac{2\pi j}{N}.
\label{eq:3.1}
\end{equation}
Unfortunately, the Fourier transform becomes unstable when the
quark number $k$ is large as we can see from Table~\ref{unstable}.
This is a well known problem; it is a consequence of the fact that
the higher Fourier components are the result of very delicate
cancellations in a a sum of alternating terms.

\begin{table}[h]
\centering
\begin{tabular}{|c|c|c|c|c|c|}
 \hline k & 3 & 6 & 9 & 12 & 15\\ \hline
 N=51  &2212.21  & 247.601 & -22.8783 & -4.53755&-0.233997  \\
 N=102 & 2212.21 & 247.601 & -22.8783 & -4.53755&-0.233997 \\
 N=204 & 2212.21 & 247.601 & -22.8783 & -4.53755&-0.233997 \\
\hline k& 18 & 21& 24 & 27 & 30  \\ \hline
 N=51  &-0.00545724& -0.0000602919 & 6.70879E-7 & 6.70879E-7&-0.0000602919  \\
 N=102 & -0.005458 & -0.0000631063 & -8.98294E-7 & 1.56917E-6&2.81435E-6 \\
 N=204 & -0.005458 & -0.0000634881 & -1.66312E-6 & -2.83726E-7&6.42123E-6 \\
 \hline
\end{tabular}
\caption{The projected determinant for a particular configuration
with different choices of $N$.
For $k$ larger than 20 the results differ for different choices of $N$,
signaling a numerical instability.
\label{unstable}}
\end{table}

For our studies using $6^3\times 4$ lattices, to investigate the phase
diagram at non-zero baryon densities, we need quark numbers of the
order $k\sim30$ which corresponds to 10 baryons. To evaluate the
projection accurately we need a better method.

\section{Winding number expansion method}

The basic idea of the new method is to use the Fourier transform of
$\log \det M(U,\phi)$ instead of the Fourier components of $\det M(U,\phi)$.
Using an approximation based on the first few components of $\log \det M(U,\phi)$
we can then analytically compute the projected determinant. The success of the
method can be traced to the fact that the Fourier components of $\log \det M(U,\phi)$
are exponentially smaller with increasing order in the expansion \footnote{This
property is also noted in a similar context in \cite{Danzer:2008xs}.}.
This is why we can approximate
the exponent very accurately with few terms which, in turn, allows us to evaluate the
Fourier components of the determinant precisely.

To see how this works we look at the hopping expansion of the $\log\det M(U,\phi)$. We
start by writing the determinant in terms of its log
\beq
\label{trlog} \det M(U,\phi)=\exp(\Tr\log M(U,\phi)).
\eeq

\begin{figure}
\centering
\includegraphics[width=11cm]{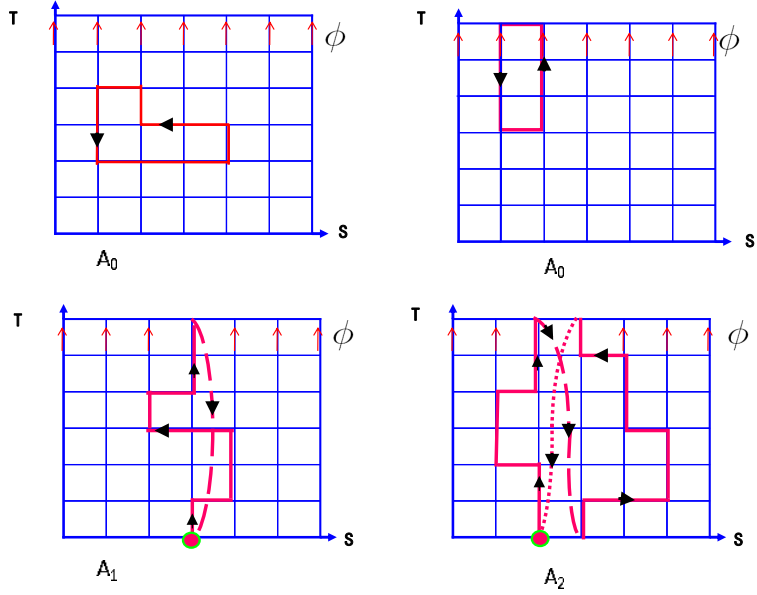}
\caption{Schematic of winding number expansion on lattice.
\label{wnem lattice}}
\end{figure}

It is well known that $\Tr\log M$ corresponds to a sum of connected quark
loops. We separate all these loops in classes in terms of the net
number of times they wrap around the lattice in the "time" direction
(see Fig.~\ref{wnem lattice}).
We have then:
\begin{eqnarray}
\label{quark loop}
&\Tr \log M(U,\phi)&=\sum_{\rm loops}L(U,\phi)\\{\nonumber}
&&= A_{0}(U)+
[\sum_{n}e^{in\phi}W_{n}(U)+e^{-in\phi}W_{n}^{\dag}(U)],
\end{eqnarray}
where $n$ is winding number of quark loops and $W_{n}$ is the weight
from the contribution of all the quark loops which have fixed
winding number $n$ in time direction. Eq.~\eqref{quark loop} can be
re-written as
\begin{eqnarray}
\label{wne expression}
&\Tr \log M(U,\phi)&= A_{0}(U)+
[\sum_{n}e^{in\phi}W_{n}(U)+e^{-in\phi}W_{n}^{\dag}(U)]\\{\nonumber}
&&=A_{0}(U)+\sum_{n} A_{n}
\cos(n\phi+\delta_{n}),
\end{eqnarray}
where $A_n \equiv 2|W_n|$ and $\delta_n \equiv \arg W_n$ are independent of $\phi$.
Using Eq.~\eqref{trlog} and Eq.~\eqref{wne expression} we get
\beq
\label{logdet wne}
\det M(U,\phi)=e^{A_{0}+A_{1}\cos(\phi+\delta_{1})+A_{2}\cos(2\phi+\delta_{2})+.....}.
\eeq
To first order in the winding number expansion we have
\beq
\det M(U,\phi)_{n=1}=e^{A_0+A_{1}\cos(\phi+\delta_{1})}.
\eeq
The Fourier transform can now be computed analytically; we have
\beq  \label{FT}
\int_{0}^{2\pi}\frac{d\phi}{2\pi}e^{-ik\phi}e^{A_0+A_{1}\cos(\phi+\delta_{1})}=e^{A_0+ik\delta_{1}}{\rm I}_k(A_1),
\eeq
where ${\rm I}_k$ is Bessel function of the first kind.

For higher orders in the winding number expansion, we compute the Fourier transform using the Taylor expansion:
\begin{eqnarray}
\label{final_wne}
&&\int_{0}^{2\pi}\frac{d\phi}{2\pi}e^{-ik\phi}e^{A_0+A_{1}\cos(\phi+\delta_{1})}e^{A_{2}\cos(2\phi+\delta_{2})+A_{3}\cos(3\phi+\delta_{3})+...}
{\nonumber}\\
&=&\int_{0}^{2\pi}\frac{d\phi}{2\pi}e^{-ik\phi}e^{A_0+A_{1}\cos(\phi+\delta_{1})}(1+A_{2}\cos(2\phi+\delta_{2})+\frac{1}{2!}A_{2}^{2}\cos(2\phi+\delta_{2})^{2}+...)\times{\nonumber}\\
&&(1+A_{3}\cos(3\phi+\delta_{3})+\frac{1}{2!}A_{3}^{2}\cos(3\phi+\delta_{3})^{2}+...)\times...{\nonumber}\\
&=&c_{00}{\mathop{\hbox{I}}}{_{k}}(A_{1})+c_{+01}{\mathop{\hbox{I}}}{_{k+1}}(A_{1})+c_{-01}{\mathop{\hbox{I}}}{_{k-1}}(A_{1})+c_{+02}{\mathop{\hbox{I}}}{_{k+2}}(A_{1})+c_{-02}{\mathop{\hbox{I}}}{_{k-2}}(A_{1})+...
\end{eqnarray}

In leading to the final result in Eq.~(\ref{final_wne}), we have used the property that
$\cos(n\phi+\delta_n) = [e^{i(n\phi+\delta_n)} + e^{-i(n\phi+\delta_n)}]/2$ and the Fourier
transform with two exponentials, according to Eq.~(\ref{FT}), gives the Bessel functions
$I_{k \pm n}(A_1)$. The coefficients $c_{00}, c_{+0n}, c_{-0n}$  are the sum of terms
for the Bessel functions $I_k(A_1), I_{k=1}(A_1), I_{k-n}(A_1)$.

Using Eq.~\eqref{final_wne} and the recursion relation for the Bessel function,
$
{\mathop{\hbox{I}}}{_{k-1}}(A)=\frac{2k}{A}{\mathop{\hbox{I}}}{_{k}}(A)+{\mathop{\hbox{I}}}{_{k+1}}(A),
$
the winding number expansion can be extended easily to higher orders.

\section{Numerical tests}

\begin{figure}
\centering
\includegraphics[width=4cm]{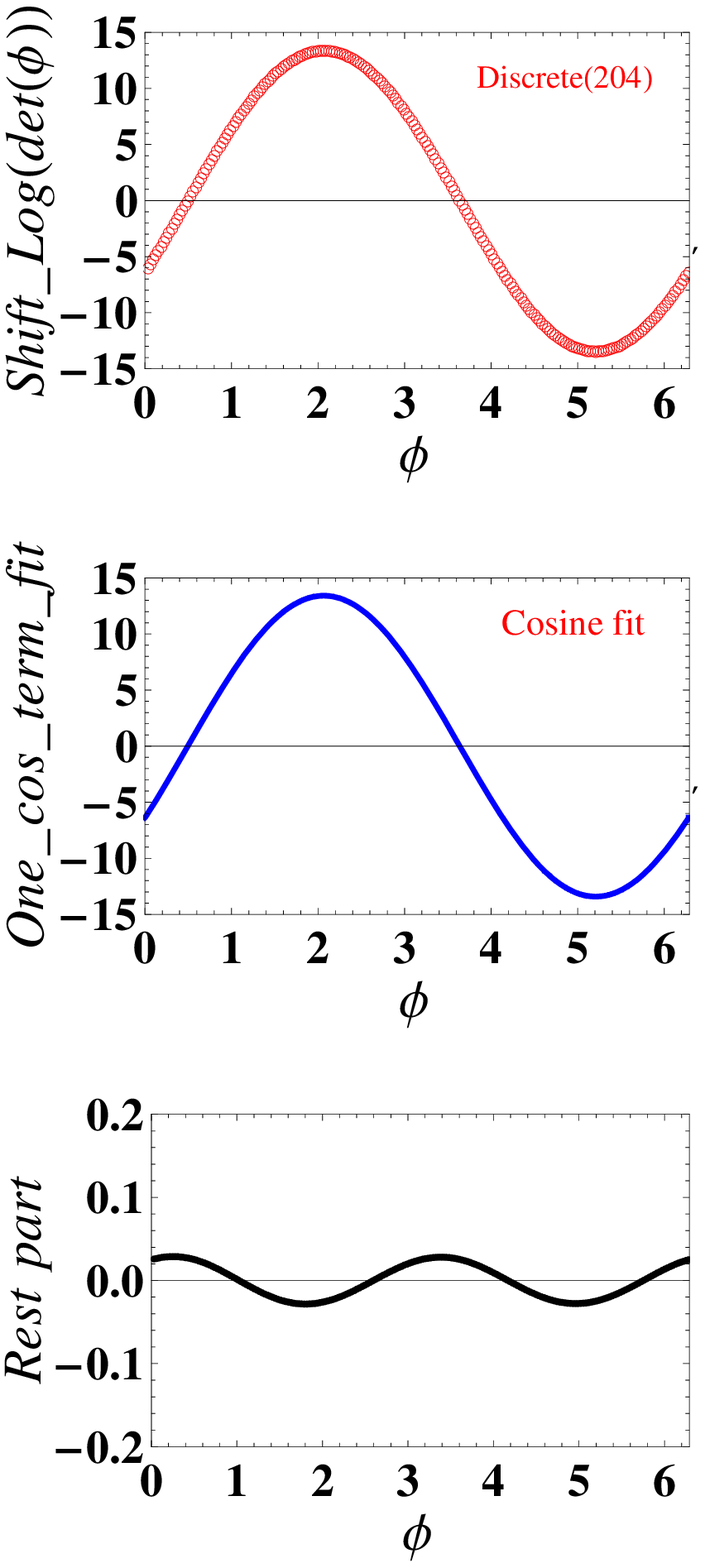}\includegraphics[width=4cm]{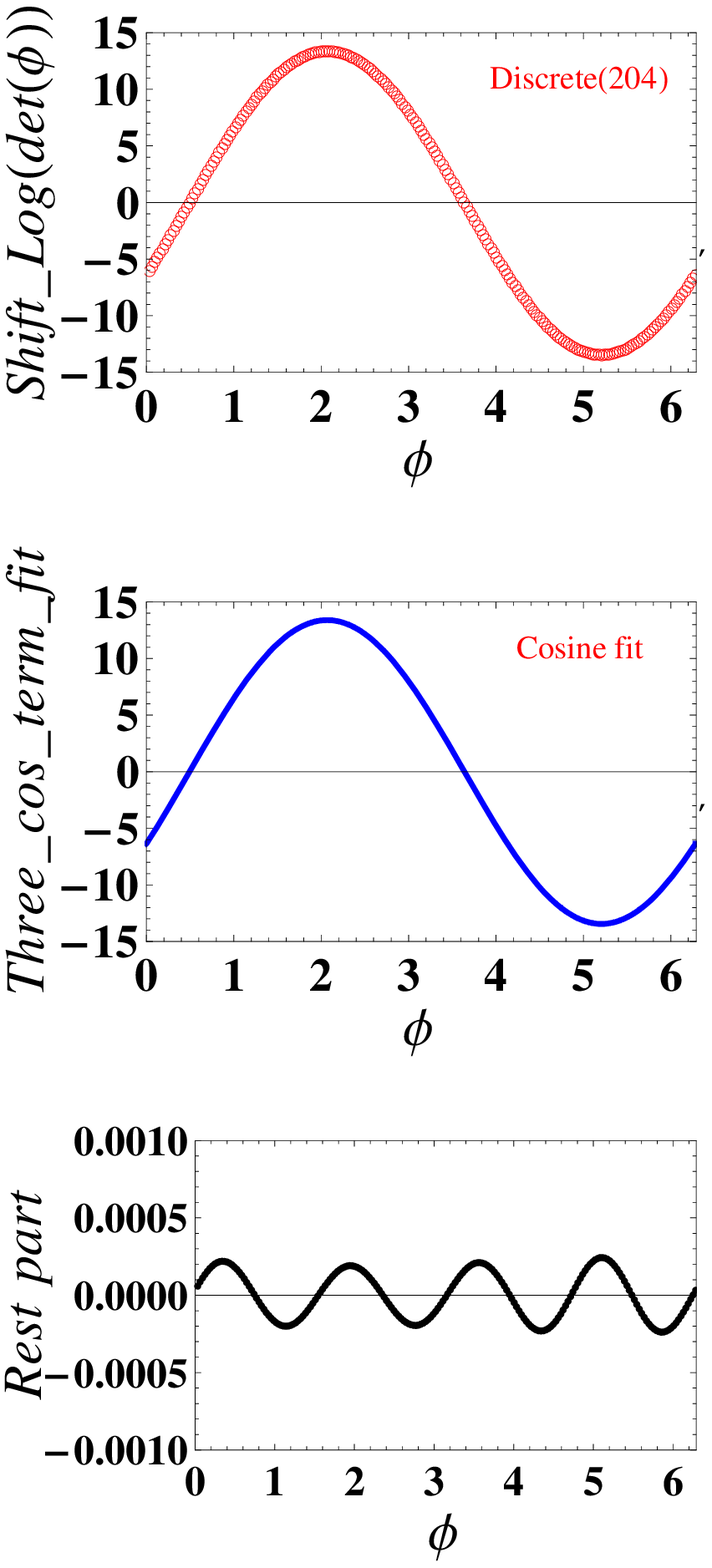}\includegraphics[width=4cm]{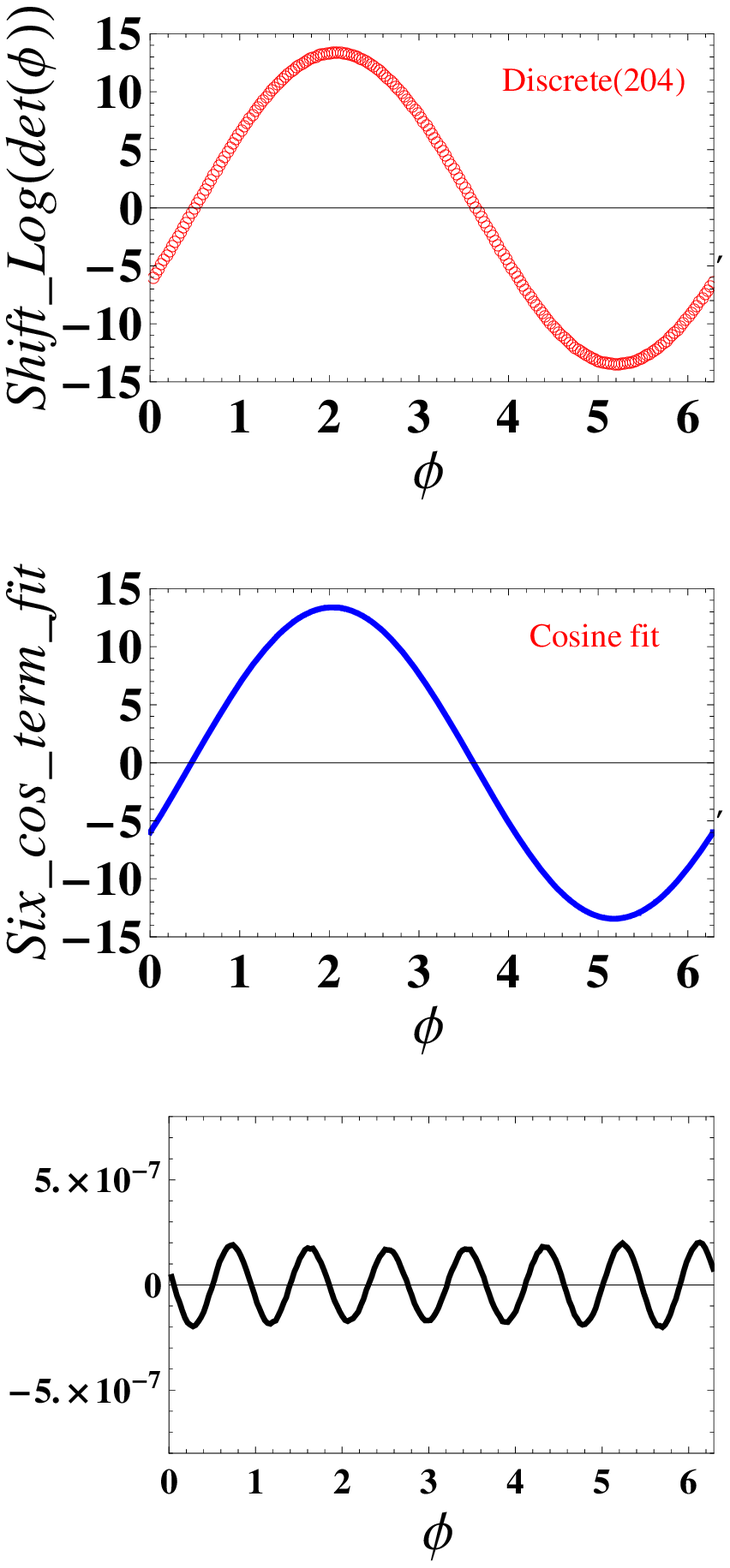}
\caption{A comparison between the exact value for $\log\det
M(U,\phi)$ and its winding number expansion to various orders. The
value of $\log\det M(U,\phi)$ is shifted so that it averages to
zero. \label{comp logdet fit}}
\end{figure}

As we mentioned before, the success of the method rests on the assumption
that we can get a very good approximation of the exponent using only a few terms in
the Fourier expansion. In Fig.~\ref{comp logdet fit}, in the top row we plot $\Tr\log M(\phi)$
evaluated at $204$ different values of $\phi$ in the interval $[0,2\pi]$ (the three
plots are the same); the second row shows an approximation using 1, 3 and
6 terms respectively. It is easy to see that the approximations are all very good --
this is due to the fact that the first term in the expansion is much larger than the
subsequent terms. To see the contributions of the higher terms in the third row,
we plot the difference between the exact value (top row) and the approximation (second row).
It is easy to see that the error of the approximation decreases very rapidly with the
number of terms. Note also that the error is well described by a cosine function: this is
due to the fact that the Fourier coefficients $A_n$ decrease exponentially with increasing $n$.
The error will be dominated by the first term that is not included in the approximation: $A_2$
for the first column, $A_4$ for the second and $A_7$ for the third.

For our tests we decided to keep a winding number expansion to the sixth order.
From Fig.~\ref{comp logdet fit} it is easy to see that the error introduced by this
approximation is of the order of $10^{-7}$ which is precise enough for our purposes. It
is important to note that to determine the coefficients $A_n$ exactly we would need
to evaluate  $\log\det M(U,\phi)$ for all possible values of $\phi$. However, we can approximate
their values using a discrete Fourier transform. To determine the right number of
points $N_d$ for the discrete $\phi_i$ needed, we compute these coefficients using an increasing number
of $N_d$. In Table~\ref{determination coefficients} we present the coefficients
determined for a particular configuration; it is easy to see that $N_d = 16$ is good enough to
determine the coefficients with sufficient precision for our purpose.

\begin{table}[h]
\centering
\begin{tabular}{|c|c|c|c|c|}
 \hline  & Nd=16 & Nd=24 & Nd=36 & Nd=200 \\ \hline
 A0  &  2.357308E+02& 2.357308E+02 & 2.357308E+02 & 2.357308E+02  \\
 A1 & -1.341400E+01 & -1.341400E+01 & -1.341400E+01 & -1.341400E+01 \\
 A2 & 2.820535E-02 & 2.820535E-02 & 2.820535E-02 & 2.820535E-02 \\
 A3  &4.135219E-04& 4.135043E-04 & 4.134942E-04 & 4.134755E-04  \\
 A4 & 2.148188E-04 & 2.147950E-04 & 2.147792E-04 & 2.147547E-04 \\
 A5 & 2.641758E-05 & 2.639153E-05 & 2.637794E-05 & 2.636227E-05 \\
 A6 & 2.289491E-06 & 2.286772E-06 & 2.291285E-06 & 2.305249E-06 \\
 \hline
\end{tabular}
\caption{The expansion coefficients determined using a discrete Fourier transform.
\label{determination coefficients}}
\end{table}

\begin{figure}
\centering
\includegraphics[width=10cm]{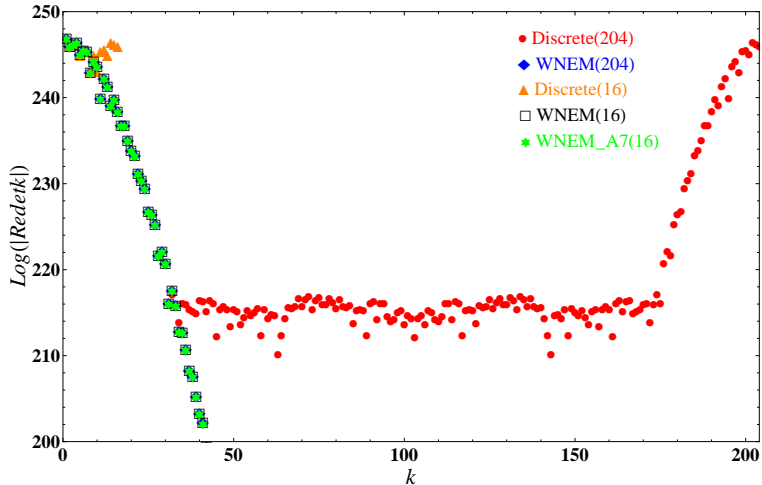}
\caption{The projected determinant as a function of the quark number. We plot here the
results of various approximation based on either the discrete Fourier transform or on
the winding number expansion.
\label{WNEM_test}}
\end{figure}

The final test, and the most important, is to determine how accurate this
approximation is when computing the projected determinant. In Fig.~\ref{WNEM_test}
we compare the result of this approximation with a high order approximation that
uses the simple discrete Fourier transform. The expectation is that as we increase
the number of quarks the value of the projected determinant decreases exponentially.
It is easy to see that this is indeed the case for all approximations at low quark number,
say for $k \le 6$.
However, as the quark number gets close to 30 the projected determinant calculated
from the old approximation (the red points) flattens out -- this is the onset of
numerical instability. The new approximation method doesn't suffer from this problem and
the projected determinant continues to decrease as the quark number is increased. We note
that the winding number expansion method (WNEM) approximates well the results from the high precision
discrete Fourier transform ($N_d = 204$) in its stability range. We conclude that the WNEM approximation
is stable and accurate for a much larger range of quark number projection.

From Fig.~\ref{WNEM_test} we also note that, as concluded above, to determine the coefficients
$A_n$ with enough precision, we don't need to use a high order approximation. Both the curves WNEM(16) and WNEM(204)
with $N_d = 16$ and 204 respectively are determined using a $6^{th}$ order WNEM. The only difference is the set of
coefficients used: WNEM(16) uses the coefficients appearing in the first column in Table~\ref{determination coefficients},
while WNEM(204) uses the coefficients in the last column. We find that the resultant projected determinants
differ only by $10^{-6}$ in relative errors for the quark number $k$ up to 50.
This is very important since it allows us to speed up the calculation tremendously without losing precision.

To compare the merits of the WNEM with the discrete Fourier transform, we compute the values of
the projected determinant calculated from the discrete Fourier transform with $N_d = 16$ only.
In Fig.\ref{determination coefficients} these results are labeled Discrete(16). It is easy to see that
this approximation is only valid up to $k=6$. This is to be compared to WNEM(16) that takes the
same computational time but can approximate the projected determinant for $k = 40$ and beyond.

One final test was to compare the $6^{th}$ order WNEM with the $7^{th}$ order WNEM. The results are
plotted in Fig.~\ref{determination coefficients}. The relative error between the two is less
than $10^{-4}$ for $k$ up to 50.  We conclude that a $6^{th}$ order
WNEM using $N_d =16$ is precise enough for the simulations for the present lattice
($6^3 \times 4$ at $\beta = 5.2$ and quark mass $\kappa=0.158$).

\section{Conclusion}

In this paper we discussed a novel approximation method for the projected determinant.
The original approximation that employs a discrete Fourier transform becomes numerically
unstable at large quark numbers which limits the region of the phase space that can be studied
using the canonical approach. We find that our method based on the winding number expansion
is numerically stable and much faster which allows us to study higher baryon numbers.

The winding number expansion method has been used successfully to carry out finite
density simulations for QCD with two and four degenerate quark flavors~ \cite{la08}. Our numerical checks
show that for all these ensembles the approximation is accurate.

Finally, we should  mention that the conclusions we have drawn in this paper are valid for
the fairly heavy quark mass we used ($m_\pi \sim 1{\rm GeV}$). As the quark mass is lowered,
we expect that more terms of $A_n$ will be needed in the expansion.




\begin{thebibliography}{99}
 \bibitem{kfl05} K.~F.~Liu,  \emph{Edinburgh 2003, QCD and Numerical Analysis Vol. III} (Springer, New York, 2005) 101, {\tt arXiv:hep-lat/0312027}.
 \bibitem{afhl05} A.~Alexandru, M.~Faber, I.~Horv\'ath, K.~F.~Liu, Phys. Rev. D {\bf 72} (2005) 114513, {\tt arXiv:hep-lat/0507020}.
 \bibitem{la07} A.~Li, A.~Alexandru, K.~F.~Liu, \pos{PoS(LAT2007)203}, {\tt arXiv:hep-lat/0711.2692}.
 \bibitem{la08} A.~Li, X.~Meng, A.~Alexandru, K.~F.~Liu, \pos{PoS(LAT2008)178},     {\tt arXiv:0810.2349v1 [hep-lat]}.
 \bibitem{Danzer:2008xs}
  J.~Danzer and C.~Gattringer, 
  {\tt arXiv:0809.2736 [hep-lat]}.

\end{thebibliography}
\end{document}